# Low-velocity-favored transition radiation

Jialin Chen[1,2,3], Ruoxi Chen[1,2], Zheng Gong[1,2], Hao Hu[4], Yi Yang[5], Xinyan Zhang[1,2], Chan Wang[1,2,6], Ido Kaminer[3,*], Hongsheng Chen[1,2,6,7,*], Baile Zhang[8,9], and Xiao Lin[1,2,3,4,*]

[1]*Interdisciplinary Center for Quantum Information, State Key Laboratory of Extreme Photonics and Instrumentation, ZJU-Hangzhou Global Scientific and Technological Innovation Center, College of Information Science & Electronic Engineering, Zhejiang University, Hangzhou 310027, China.*

[2]*International Joint Innovation Center, the Electromagnetics Academy at Zhejiang University, Zhejiang University, Haining 314400, China.*

[3]*Department of Electrical and Computer Engineering, Technion-Israel Institute of Technology, Haifa 32000, Israel.*

[4]*School of Electrical and Electronic Engineering, Nanyang Technological University, Singapore 639798, Singapore.*

[5]*Department of Physics, University of Hong Kong, Hong Kong 999077, China.*

[6]*Key Lab of Advanced Micro/Nano Electronic Devices & Smart Systems of Zhejiang, Jinhua Institute of Zhejiang University, Zhejiang University, Jinhua 321099, China.*

[7]*Shaoxing Institute of Zhejiang University, Zhejiang University, Shaoxing 312000, China.*

[8]*Division of Physics and Applied Physics, School of Physical and Mathematical Sciences, Nanyang Technological University, Singapore 637371, Singapore.*

[9]*Centre for Disruptive Photonic Technologies, Nanyang Technological University, Singapore 637371, Singapore.*

[*]*Corresponding authors. Email: xiaolinzju@zju.edu.cn (X. Lin); kaminer@technion.ac.il (I. Kaminer); hansomchen@zju.edu.cn (H. Chen)*

**When a charged particle penetrates through an optical interface, photon emissions emerge — a phenomenon known as transition radiation. Being paramount to fundamental physics, transition radiation has enabled many applications from high-energy particle identification to novel light sources. A rule of thumb in transition radiation is that the radiation intensity generally decreases with the particle velocity $v$; as a result, low-energy particles are not favored in practice. Here we find that there exist situations where transition radiation from particles with extremely low velocities (e.g. $v/c < 10^{-3}$) exhibits comparable intensity as that from high-energy particles (e.g. $v/c = 0.999$), where $c$ is light speed in free space. The comparable radiation intensity implies an extremely high photon extraction efficiency from low-energy particles, up to eight orders of magnitude larger than that from high-energy particles. This exotic phenomenon of low-velocity-favored transition radiation originates from the excitation of Ferrell-Berreman modes in epsilon-near-zero materials. Our findings may provide a promising route towards the design of integrated light sources based on low-energy electrons and specialized detectors for beyond-standard-model particles.**

Transition radiation occurs whenever a charged particle moves across an inhomogeneous region [1-8]; as shown in Fig. 1. One unique feature of transition radiation is that its radiation intensity is linearly proportional to the Lorentz factor $\gamma = (1 - v^2/c^2)^{-1/2}$ [9], if the particle velocity $v$ approaches the light speed $c$ in free space. This feature lays the foundation for many applications, including transition radiation detectors [10-12], useful for the identification of particles with extremely high momenta (e.g. $P > 100 \text{ GeV}/c$ or $\gamma > 10^5$) [9], as well as advanced light sources at the terahertz, ultraviolet, and X-ray regimes [13-19]. However, all these transition-radiation-based devices rely on high-energy particles, whose generation requires a giant and complex acceleration infrastructure and thus hinders the enticing on-chip applications of transition radiation.

Another feature of transition radiation is that its occurrence has no specific requirements on the particle velocity [1-8]. This feature is distinct from Cherenkov radiation [20-25], which occurs when a charged particle moves inside a homogeneous material with a velocity exceeding the phase velocity of light, namely the Cherenkov threshold [26,27]. As such, all applications of Cherenkov radiation are limited by the Cherenkov threshold, despite its enormous applications [28-34], including particle detectors, light sources, imaging, and photodynamic therapy. By contrast, transition radiation is applicable to develop novel light sources without fundamental restrictions on the particle velocities. However, the application of transition radiation based on low-energy particles remains largely unexplored. One reason is that the transition-radiation intensity generally decreases when the particle velocity decreases. For example, the intensity of conventional transition radiation from low-energy particles with $v/c = 0.1$ can be two orders of magnitude weaker than that from high-energy particles with $v/c = 0.9$, as exemplified in Fig. 2. The enhancement of transition radiation from low-energy particles remains an open challenge in science and technology.

Here we reveal a feasible route to enhance transition radiation from low-energy particles (e.g. free electrons) by exploiting the Ferrell-Berreman mode, which was first identified by Ferrell in 1958 in metal films at the ultraviolet regime [35] and later separately discussed by Berreman in 1963 in cubic ionic films at the mid-infrared regime [36]. This mode is intrinsically radiative and appears near the frequency at which the relative permittivity of materials approaches zero. Moreover, the Ferrell-Berreman mode has enabled many applications [37-40] from imaging, sensing to thin-film characterization. The transition radiation of Ferrell-Berreman modes has also been extensively studied since 1958 [4,41-44]. However, among these studies, the intensity dependence of the emitted Ferrell-Berreman mode on the particle velocity has been rarely discussed so far. Moreover, whether the Ferrell-Berreman mode can largely enhance the transition radiation from low-energy particles remains unknown.

Here we find that due to the excitation of Ferrell-Berreman modes, low-energy particles with an extremely low velocity (e.g. $v/c < 10^{-3}$) could emit equally strong transition radiation as high-energy particles ($v/c = 0.999$). Consequently, the photon extraction efficiency from low-energy particles could be eight orders of magnitude larger than that from high-energy particles, while the intensity of transition radiation from these particles is the same. This exotic phenomenon of free-electron radiation is then denoted as low-velocity-favored transition radiation, which is in a similar rationale but fundamentally different from low-velocity-favored Smith-Purcell radiation [45]. The revealed low-velocity-favored transition radiation indicates a promising route to enhance the particle-matter interaction, which may be exploited to design specialized detectors for beyond-standard-model particles with extremely-low kinetic energy (e.g. detection of unknown millicharged dark matter [46-48]) and integrated light sources from low-energy electrons.

We begin with the introduction of transition radiation; see derivation in supplementary section S1. An electron moves along the $+\hat{z}$ direction and perpendicularly penetrates through a thin epsilon-near-zero slab with a thickness $d$ [Fig. 1]. The epsilon-near-zero slab (namely region 2), for example, is constructed by hexagonal boron nitride (BN) [49-55] with a relative permittivity of $[\varepsilon_\perp, \varepsilon_\perp, \varepsilon_z]$, which has $\varepsilon_z \to 0$ around 24.5 THz. Both the superstrate (region 1) and the substrate (region 3) are free space with a relative permittivity of $\varepsilon_1 = \varepsilon_3 = 1$.

Within the framework of macroscopic Maxwell equations, the induced radiation fields in regions 1 and 3 can be calculated, and the total angular spectral energy density of transition radiation is obtained as $U(\omega, \theta) = U_1(\omega, \theta) + U_3(\omega, \theta)$, where $U_1(\omega, \theta)$ and $U_3(\omega, \theta)$ are the angular spectral energy densities of backward and forward radiation, respectively, and $\theta$ is the radiation angle. Accordingly, the total energy spectrum can be expressed as $W(\omega) = \int_0^{\pi/2} U(\omega, \theta) \cdot (2\pi \sin\theta) d\theta$.

As shown in Fig. 2a, $W(\omega)$ is not only a function of the particle velocity but also sensitive to the frequency, due to the dispersive nature of BN. The BN thickness is 1 nm (see the influence of thickness on low-velocity favored transition radiation in Fig. S3); alternatively, one may drill a hole at the slab along the electron trajectory to avoid the potential scattering of swift electrons. Fig. 2a shows that the frequency spectral feature of transition radiation near the frequency with $\varepsilon_z \to 0$ is different from the other frequency regimes. For better illustration, Fig. 2b shows $W(\omega)$ as a function of the electron velocity at three representative frequencies, namely 24.5 THz (within the first reststrahlen band of BN) with $\varepsilon_\perp = 7.7 + 0.01i$ and $\varepsilon_z = -0.05 + 0.04i$, 42 THz (within the second reststrahlen band) with $\varepsilon_\perp = -34.8 + 4.6i$ and $\varepsilon_z = 2.7 + 0.0005i$, and 35 THz (outside these two reststrahlen bands) with $\varepsilon_\perp = 11.6 + 0.1i$ and $\varepsilon_z = 2.5 + 0.001i$. At 35 THz or 42 THz without $\varepsilon_z \to 0$, $W(\omega)$ monotonically increases with $v$ in Fig. 2b. By contrast, at $\omega_0/2\pi = 24.5$ THz with $\varepsilon_z \to 0$, $W(\omega_0)$ first increases with $v$ if $v < v_A$, becomes insensitive to the variation of $v$ if $v \in [v_A, v_B]$, then decreases with $v$ if $v \in [v_B, v_C]$, and increases again with $v$ if $v > v_C$, where $v_A = 0.4 \times 10^{-3}c$, $v_B = 2.9 \times 10^{-1}c$, and $v_C = 0.8c$. Remarkably, Fig. 2b also shows that the radiation intensity at the frequency with $\varepsilon_z \to 0$ can be two orders of magnitude larger than that at the frequency without $\varepsilon_z \to 0$ if $v/c < 10^{-1}$. Moreover, the values of $W(\omega_0)$ are the same, if the electron velocity is equal to $v_A$, $v_B$ or $v_D$, where $v_D = 0.999c$. For these velocities, the corresponding kinetic energy varies from $E_{k,A} = 40.9$ meV, $E_{k,B} = 22.9$ keV to $E_{k,D} = 10.9$ MeV. This exotic feature at 24.5 THz in Fig. 2a&b indicates that the transition radiation from low-energy particles can achieve the same radiation intensity as that from high-energy particles.

According to Fig. 2a&b, the photon extraction efficiency is further obtained as $\eta(\omega) = \frac{W(\omega)}{\hbar\omega \cdot E_k}$, where $E_k = m_e c^2 (\gamma - 1)$ is the kinetic energy, $m_e$ is the rest mass, and $\gamma = (1 - v^2/c^2)^{-1/2}$ is the Lorentz factor. Figure 2c shows the photon extraction efficiency as a function of frequency and velocity. From Fig. 2c, if $v/c < 10^{-2}$, the maximum value of $\eta(\omega)$ always appears near the frequency with $\varepsilon_z \to 0$. Upon close inspection, Fig. 2d shows $\eta(\omega)$ as a function of velocity at three representative frequencies. When $v/c < 10^{-2}$, the photon extraction efficiency at 24.5 THz is nearly three orders of magnitude greater than those at 42 THz and 35 THz. Moreover, at $\omega_0/2\pi = 24.5$ THz, the value of $\eta(\omega_0)$ at point A with $v_A/c = 0.4 \times 10^{-3}$ is eight orders of magnitude higher than that at point D with $v_D/c = 0.999$ in Fig. 2d, although the values of $W(\omega_0)$ at these two points are the same in Fig. 2b.

Besides the unique frequency spectral features in Fig. 2, this low-velocity-favored transition radiation also has exotic angular spectral features in Fig. 3. Figure 3a&d illustrates the angular spectral energy density $U(\omega, \theta)$ as a function of velocity and radiation angle $\theta$. At the frequency with $\varepsilon_z \to 0$, the dependence of $U(\omega, \theta)$ on $v$ is not monotonical but rather complex in Fig. 3a. Moreover, if the electron velocity is small (e.g. $v/c < 0.3$), the maximum of $U(\omega, \theta)$ still appears at a relatively-large radiation angle ($> 80°$) with a relatively-large angular width ($> 50°$) in Fig. 3a. This feature for low-energy electrons is different from that for high-energy electrons. Generally, if $v \to c$ or $\gamma \gg 1$, the maximum of $U(\omega, \theta)$ would appear at $\theta \to 0°$ with a very narrow angular width ($< 0.1°$) [9]. Besides, the angular spectral feature of transition radiation at the frequency with $\varepsilon_z \to 0$ in Fig. 3a is entirely different from those at frequencies without $\varepsilon_z \to 0$ in Fig. 3d. At the frequency without $\varepsilon_z \to 0$, $U(\omega, \theta)$ in Fig. 3d monotonically increases with $v$. Meanwhile, the maximum of $U(\omega, \theta)$ starts to appear at a relatively-small radiation angle ($< 45°$) if $v/c > 0.3$ in Fig. 3d.

To confirm the unique angular feature above, Fig. 3b-c shows the field distribution of the excited waves at the frequency with $\varepsilon_z \to 0$. Remarkably, the field strength with $v/c = 0.001$ in Fig. 3b is comparable to that with $v/c = 0.999$ in Fig. 3c. However, the excited waves mainly propagate to the directions almost parallel to the interface (with $\theta \to 90°$) if $v/c = 0.001$ in Fig. 3b, while most of the excited waves would propagate to the direction almost perpendicularly to the interface (with $\theta \to 0°$) if $v/c = 0.999$ in Fig. 3c. For comparison, we show the excited waves at the frequency without $\varepsilon_z \to 0$ (e.g. in Fig. 3e-f & Fig. S2). By contrast, the field strength with $v/c = 0.001$ in Fig. 3e is much weaker than that with $v/c = 0.999$ in Fig. 3f.

Since the low-velocity-favored transition radiation mainly occurs near the frequency with $\varepsilon_z \to 0$, its origin should be closely related to the excitation of Ferrell-Berreman modes [35,36]. This mode itself has been extensively studied, including that in anisotropic systems [36,56-58]. The transition radiation of Ferrell-Berreman modes has also been extensively discussed but is focused on the isotropic materials (e.g. a metal slab) [4,35,41-44]. While its study in other complex anisotropic systems, including uniaxial materials (e.g. BN), has been relatively-less explored. For uniaxial materials, $\varepsilon_z \to 0$ and $\varepsilon_\perp \to 0$ are in principle both possible. It is then natural to ask whether $\varepsilon_z \to 0$ or $\varepsilon_\perp \to 0$ is the crucial parameter to create the low-velocity-favored transition radiation.

To address this issue, Fig. 4a re-plots the radiation spectrum of transition radiation from BN in Fig. 2a under three fixed velocities. For BN, we have $\varepsilon_\perp \to 0$ near 48.2 THz (namely $\varepsilon_\perp = 0.02 + 0.08i$ and $\varepsilon_z = 2.8 + 0.0003i$), in addition to $\varepsilon_z \to 0$ near 24.5 THz. A radiation peak, which is a characteristic signature of the Ferrell-Berreman mode, always shows up near 24.5 THz but does not emerge near 48.2 THz in Fig. 4a. Hence, we can argue that $\varepsilon_z \to 0$, instead of $\varepsilon_\perp \to 0$, plays a crucial role in the excitation of Ferrell-Berreman modes. Moreover, Fig. 4b shows the transition radiation from various uniaxial epsilon-near-zero materials with $|\varepsilon_z| \to 0$, where the material loss is neglected and the other structural setup is the same as that in Fig. 2a. The phenomenon of the low-velocity-favored transition radiation always appears, no matter $\text{Re}(\varepsilon_z) > 0$ or $\text{Re}(\varepsilon_z) < 0$. Meanwhile, the appearance of this phenomenon is relatively insensitive to the values of $\varepsilon_\perp$ in Fig. 4b. From Fig. 4a-b, we then conclude that $|\varepsilon_z| \to 0$ plays a determinant role in the creation of the low-velocity-favored transition radiation. Therefore, it is better to exploit $|\varepsilon_z| \to 0$, instead of $|\varepsilon_\perp| \to 0$, for the design of novel light sources based on low-energy electrons.

In conclusion, we have demonstrated a feasible route to achieve the low-velocity-favored transition radiation by exploiting the Ferrell-Berreman mode in epsilon-near-zero materials. Such an exotic phenomenon of free-electron radiation can simultaneously achieve strong emission intensity and high photon extraction efficiency readily from low-energy electrons. Due to the abundance of epsilon-near-zero materials in nature or through judicious nanofabrication [59-63], our finding of low-velocity-favored transition radiation can apply to a broad range of frequencies, e.g., terahertz, mid-infrared and ultraviolet. Therefore, our finding not only enriches the physics of free-electron radiation but also broadens potential applications of Ferrell-Berreman modes, especially for the design of integrated free-electron light sources that are highly efficient and tunable.

**Acknowledgement**
X.L acknowledges the support partly from the National Natural Science Fund for Excellent Young Scientists Fund Program (Overseas) of China, the National Natural Science Foundation of China under Grant No. 62175212, Zhejiang Provincial Natural Science Fund Key Project under Grant No. LZ23F050003, the Fundamental Research Funds for the Central Universities (2021FZZX001-19), and Zhejiang University Global Partnership Fund. H.C. acknowledges the support from the Key Research and Development Program of the Ministry of Science and Technology under Grants No. 2022YFA1404704, 2022YFA1404902, and 2022YFA1405200, the National Natural Science Foundation of China under Grants No.11961141010 and No. 61975176. J.C. acknowledges the support from the Chinese Scholarship Council (CSC No. 202206320287). I.K. acknowledges the support from the Israel Science Foundation under Grant No. 3334/19 and the Israel Science Foundation under Grant No. 830/19. Y.Y. acknowledges the support from the start-up fund of the University of Hong Kong and the National Natural Science Foundation of China Excellent Young Scientists Fund (HKU 12222417). B.Z. acknowledges the support from Singapore National Research Foundation Competitive Research Program No. NRF-CRP23-2019-0007.

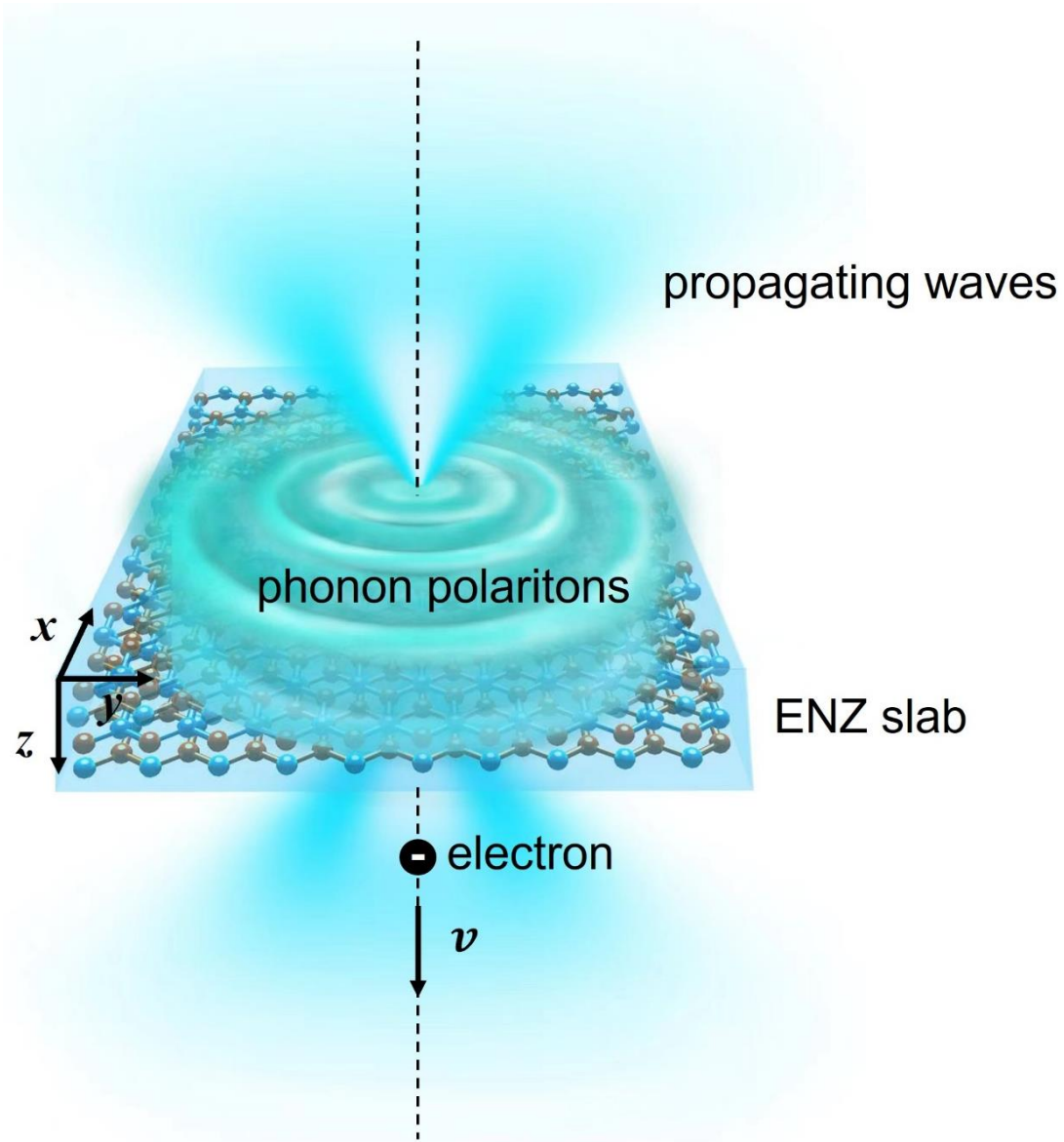

**Figure 1. Schematic of low-velocity-favored transition radiation from a uniaxial epsilon-near-zero (ENZ) material.** A swift electron perpendicularly penetrates through a BN slab with a relative permittivity of $[\varepsilon_\perp, \varepsilon_\perp, \varepsilon_z]$, where $|\varepsilon_z| \to 0$ near 24.5 THz. While both the propagating waves and phonon polaritons would be excited, below we focus on the light emission in the far field.

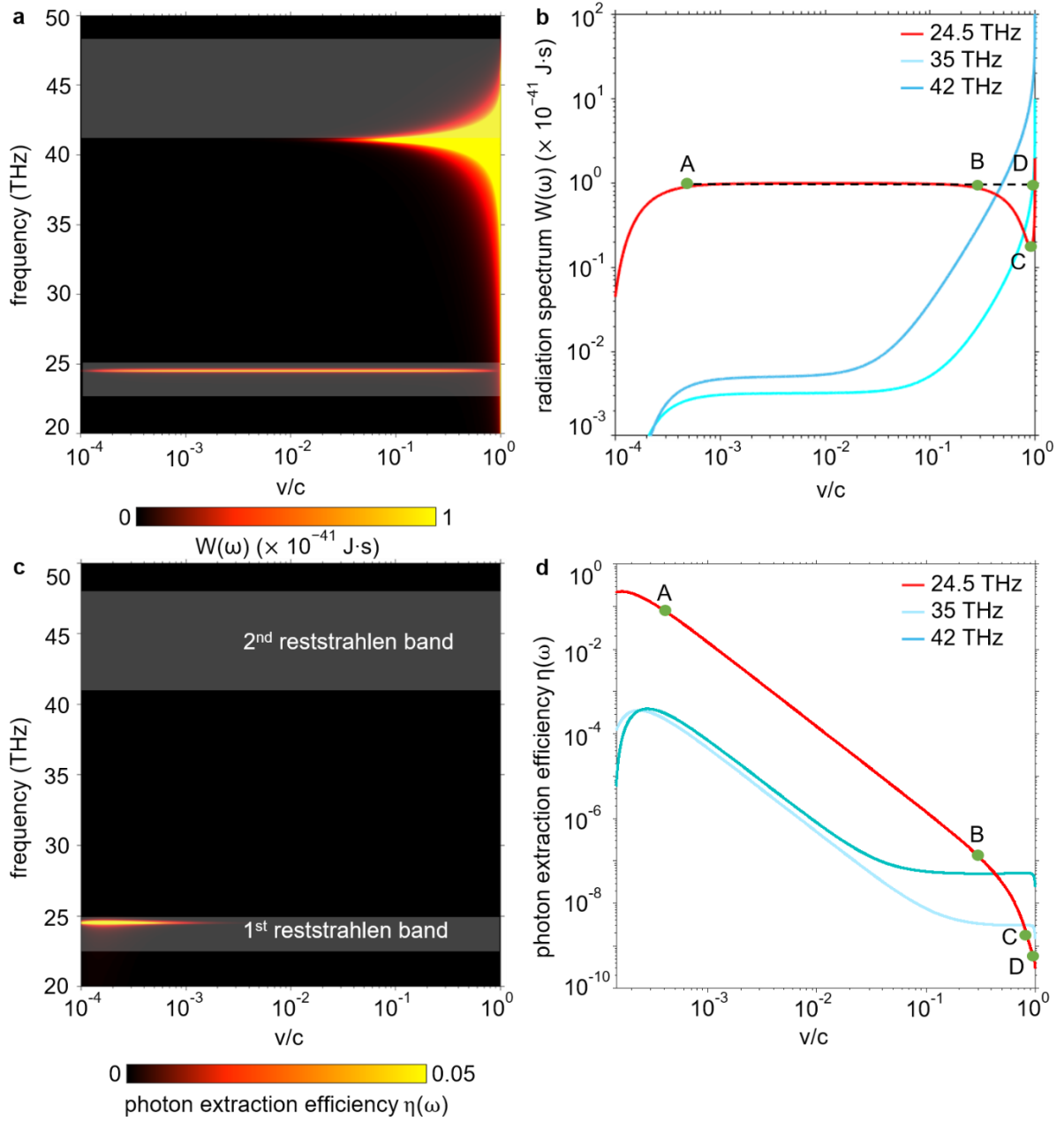

**Figure 2. Frequency spectral feature of low-velocity-favored transition radiation.** (**a**) Radiation spectrum $W(\omega)$ of the excited propagating waves as a function of the electron velocity $v$ and the frequency. (**b**) Radiation spectrum as a function of $v$ at three representative frequencies. (**c,d**) Photon extraction efficiency $\eta(\omega) = \frac{W(\omega)}{\hbar\omega \cdot E_k}$, where $E_k$ is the kinetic energy.

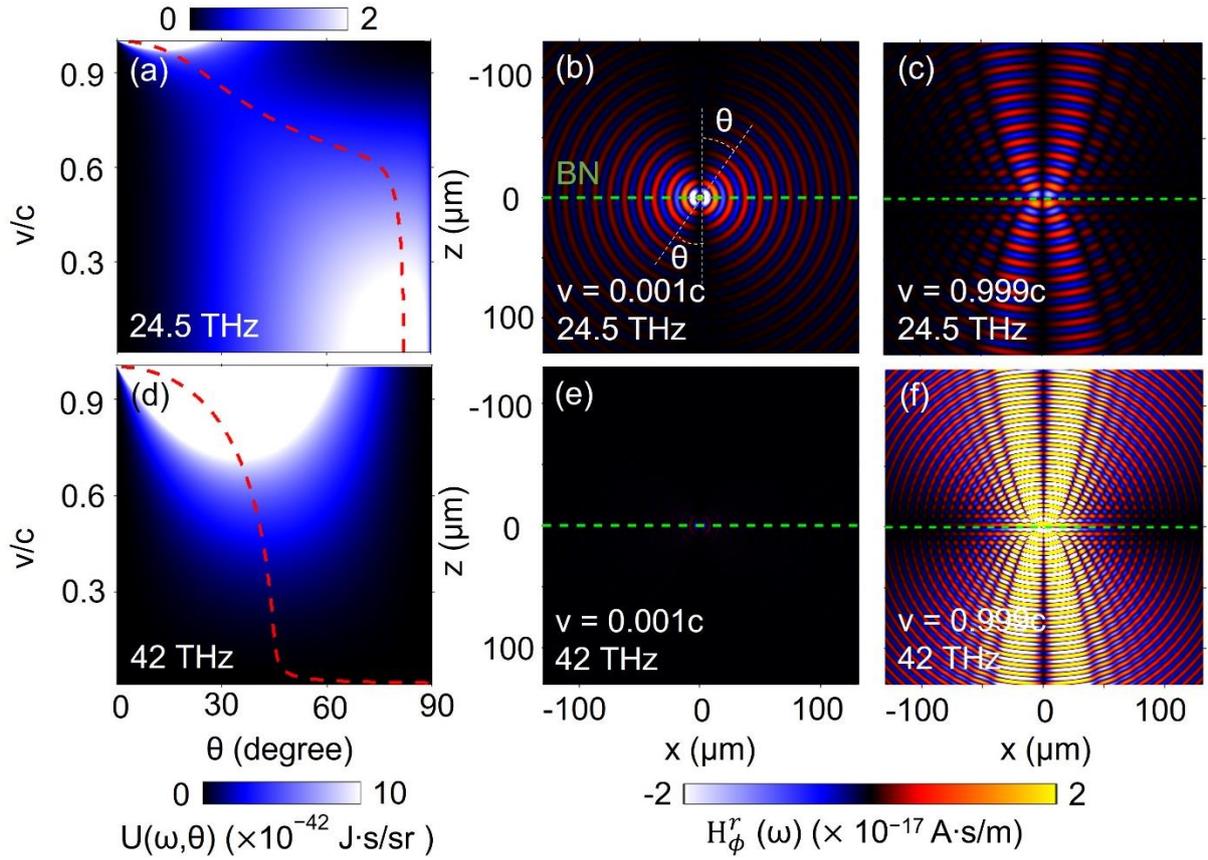

**Figure 3. Angular spectral feature of low-velocity-favored transition radiation.** (**a, d**) Angular spectral energy density $U(\omega, \theta)$ of transition radiation as a function of the radiation angle $\theta$ and the electron velocity $v$. Each red dashed line indicates the angular trajectory of the maximum of $U(\omega, \theta)$. (**b-c, e-f**) Distribution of the excited magnetic field $H_\phi^r$, where each green line represents the BN slab.

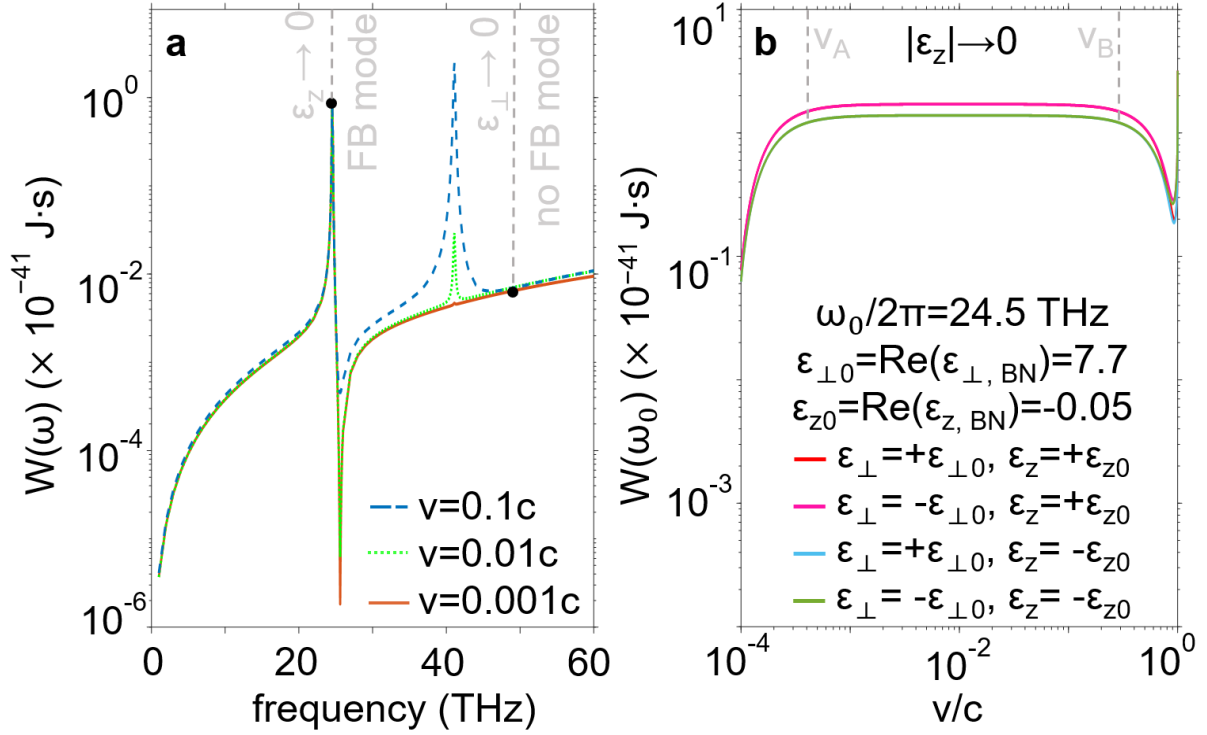

**Figure 4. Existence of the low-velocity-favored transition radiation in various epsilon-near-zero materials.** (**a**) Radiation spectrum of transition radiation from a uniaxial BN slab. The radiation peak near the frequency with $\varepsilon_z \to 0$ is related to the excitation of Ferrell-Berreman (FB) modes, while there is no radiation peak near the frequency with $\varepsilon_\perp \to 0$. (**b**) Radiation spectrum of transition radiation from uniaxial materials with $|\varepsilon_z| \to 0$; the structural setup is the same as Fig. 2a, except for the permittivity of region 2.

13